\documentclass[8.5pt,twoside,twocolumn]{article}
\usepackage[super,sort&compress,comma]{natbib} 
\usepackage[version=3]{mhchem}
\usepackage{times,mathptm}
\usepackage{sectsty}
\usepackage{balance} 
\usepackage{color}

\usepackage{graphicx} 
\usepackage{lastpage}
\usepackage[format=plain,justification=raggedright,singlelinecheck=false,font=small,labelfont=bf,labelsep=space]{caption} 
\usepackage{fancyhdr}
\usepackage{epic}
\usepackage{color}

\definecolor{red}{rgb}{1,0,0}
\definecolor{blue}{rgb}{0,0,1}
\definecolor{black}{rgb}{0,0,0}

\usepackage{url}

\begin{document}

\thispagestyle{plain}
\fancypagestyle{plain}{
\renewcommand{\headrulewidth}{1pt}}
\renewcommand{\thefootnote}{\fnsymbol{footnote}}
\renewcommand\footnoterule{\vspace*{1pt}%
\hrule width 3.4in height 0.4pt \vspace*{5pt}} 

\makeatletter 
\renewcommand\@biblabel[1]{#1}            
\renewcommand\@makefntext[1]%
{\noindent\makebox[0pt][r]{\@thefnmark\,}#1}
\makeatother 
\renewcommand{\figurename}{\small{Fig.}~}
\sectionfont{\large}
\subsectionfont{\normalsize} 

\fancyfoot{}
\fancyhead{}
\renewcommand{\headrulewidth}{1pt} 
\renewcommand{\footrulewidth}{1pt}
\setlength{\arrayrulewidth}{1pt}
\setlength{\columnsep}{6.5mm}
\setlength\bibsep{1pt}

\twocolumn[
  \begin{@twocolumnfalse}
\noindent\LARGE{\textbf{Testing the Foundations of Classical Entropy: Colloid Experiments}}
\vspace{0.6cm}

\noindent\large{\textbf{Michael E. Cates{$^{\ast}$}$^{a,b}$, Vinothan N. Manoharan$^{c,d}$}}


\vspace{0.6cm}

\noindent \normalsize{
Defining the entropy of classical particles raises a number of paradoxes and ambiguities, some of which have been known for over a century. Several, such as Gibbs' paradox, involve the fact that classical particles are distinguishable, and in textbooks these are often `resolved' by appeal to the quantum-mechanical indistinguishability of atoms or molecules of the same type. However, questions then remain of how to correctly define the entropy of large poly-atomic particles such as colloids in suspension, of which no two are exactly alike. By performing experiments on such colloids, one can establish that certain definitions of the classical entropy fit the data, while others in the literature do not. Specifically, the experimental facts point firmly to an `informatic' interpretation that dates back to Gibbs: entropy is determined by the number of microstates that we as observers {\em choose to treat as equivalent} when we identify a macrostate. This approach, unlike some others, can account for the existence of colloidal crystals, and for the observed abundances of colloidal clusters of different shapes. We also address some lesser-known paradoxes whereby the physics of colloidal assemblies, which ought to be purely classical, seems to involve quantum mechanics directly. The experimental symptoms of such involvement are predicted to be `isotope effects' in which colloids with different inertial masses, but otherwise identical sizes and properties, show different aggregation statistics. These paradoxes are caused by focussing one's attention on some classical degrees while neglecting others; when all are treated equally, all isotope effects are found to vanish.}

\vspace{0.5cm}
 \end{@twocolumnfalse}
  ]

\footnotetext{\textit{$^{a}$~SUPA, School of Physics and Astronomy, University of Edinburgh, Mayfield Road, Edinburgh EH9 3JZ, UK; $^{b}$~DAMTP, Centre for Mathematical Sciences, University of Cambridge, Cambridge CB3 OWA, UK; $^{c}$~Harvard John A. Paulson School of Engineering and Applied Sciences, Harvard University, Cambridge, MA 02138 USA; $^{d}$~Department of Physics, Harvard University, Cambridge, MA 02138 USA.}}

\section{Introduction}

Understanding the phase behaviour of colloidal suspensions has been one of the crowning successes of classical statistical mechanics. Their various equilibrium phases\cite{lekkerkerker_colloids_2011} can be predicted by computing the Helmholtz free energy $F(T,V,N) = E-TS$ for $N$ particles in a volume $V$ at temperature $T$. 

For hard spheres, the interaction energy $E$ is zero and the entropy, $S$, is the sole determinant of the phase diagram. In many colloids the suspended particles have a hard core, a steric or electrostatic repulsion of short range compared to the size of the core, and negligible interaction at larger distances. These indeed behave like hard spheres and their properties confirm all aspects of the theoretical phase diagram for such spheres\cite{poon_colloids_2004}. Adding attractions at larger distances complicates things (since $E$ and $S$ both matter) but the calculations can still be done, and agreement is just as impressive\cite{lekkerkerker_colloids_2011}.

There are two puzzling niggles with this success. First: what happened to the solvent? The calculations are performed for particles with hard sphere (and/or other) interactions, as though they resided in a vacuum. Is it really true that the solvent can be ignored altogether, once the effective interactions between colloidal particles (which of course may depend on what the solvent is) have been worked out?
Secondly, a colloidal suspension may contain, in a typical sample, of order $10^{12}$ nearly spherical particles, each a micron in diameter. Each of these particles contains in turn around $10^{12}$ atoms. It follows that, if we allow for small variations in shape as well as in size, {\em no two colloidal particles are exactly alike}. Despite this, the phase diagrams observed for colloidal systems are the same as those calculated by statistical mechanics on the basis that the interacting particles concerned are {\em indistinguishable}. 

We address the issue of the `missing solvent' later in this article, but focus first on the apparent role of indistinguishability. Recall that this is a fundamentally quantum-mechanical concept, and applies only to particles that are molecularly identical. This does not hold for colloidal suspensions, for the reasons just given. Sethna has coined the term `undistinguished' particles for classical systems like these in which particles could be distinguished, but in practice are not\cite{sethna_statistical_2006}. It seems deeply reasonable that undistinguished and indistinguishable particles should have the same thermodynamic behaviour, at least so long as one remains firmly in the domain of classical physics. Within this domain, fermions and bosons have the same statistics, and although factors of $\hbar$ might enter a calculation, they should all cancel out in the end. Colloids are large enough to be classical in this sense.

Such an equivalence seems particularly reasonable if one takes an `informatic' definition\cite{rosenkrantz_e._1983} of the entropy $S$. In this informatic view, $S = k_B \ln W$ where $W$ is the number of distinct and equiprobable microstates that we choose to treat as equivalent when defining a macrostate. (Slightly more generally, $S = -k_B\sum p_i\ln p_i$ where $p_i$ is the probability of microstate $i$ as determined within an ensemble of our choosing.) It follows that if we {\em choose} to make no distinction between colloidal particles---an attractive proposition for the experimentalist, who would likely prefer not to label each of the $10^{12}$ individual particles she can follow in her microscope---then all our calculations become the same as if there actually is no distinction. (In technical terms, the classical partition function for $N$ labelled particles must always be divided by a factor $N!$ to account for permutations among them\cite{mcquarrie_statistical_1976,frenkel_introduction_2006}: this holds whether the particles are actually indistinguishable or merely `undistiguished'.) The informatic view thus neatly explains why the statistical mechanics calculations capture the observed phase behaviour of colloidal suspensions.

Intriguingly however, the equivalence of indistinguishable and undistinguished particles is not accepted by all interpretations of classical statistical mechanics. In particular there is one school of thought, most clearly elucidated in an influential textbook by the distinguished physicist Shang-Keng Ma\cite{ma_statistical_1985}, that asserts that for the entropy of a system to be objectively real, it must be a dynamical quantity and not an informatic one. According to this `kinetic' approach, $S = k_B \ln W$ where $W$ measures the volume in phase space that the system can explore on the time scale of an experiment. The precise definition of this time-scale dependent volume is problematic, as discussed carefully by Ma. (Specifically, only a tiny fraction of accessible states are ever sampled in practice.) Nonetheless one can sympathise with his view that the entropy ought to be a property of a thermodynamic system alone, and hence definable without reference to the informatic state of an observer. The `kinetic' view of entropy has carried significant weight, for instance in the community working on glasses\cite{glassMa}.

The informatic and kinetic views of entropy would be in harmony if they always gave equivalent predictions for macroscopic behaviour such as phase equilibria. However, they do not; and wherever this issue has been looked at carefully\cite{eastman_entropy_1933, goldstein_reality_2008,speedy_relations_1999}, the informatic view has been found correct, and the kinetic one found wanting.

This point is best illustrated by a colloidal experiment that has been done innumerable times in laboratories around the world. If a suspension of monodisperse hard colloidal spheres is prepared in a homogeneous fluid state at (say) a volume fraction $\phi = 56\%$, this state will crystallize. (In the phase diagram, this initial state lies in the metastable fluid but has a density below the glass transition. Crystallization is therefore rapid, and the identity of the equilibrium state as a colloidal crystal is unambiguous.) In the fluid, the spheres can easily swap places whereas in the crystal, they cannot. It is only a slight oversimplification to say that the diffusivity of the colloidal particles in the crystal is negligible, so that particle swaps are entirely absent on experimental time scales\footnote{Of course, colloid diffusivity is not entirely negligible in the crystal, nor is it so fast in the liquid as to explore all states on an experimental timescale. But nonetheless, the liquid samples vastly more permutations than the crystal on experimental timescales.}. For indistinguishable particles, the entropy {\em gain} on transforming from liquid to crystal is extensive, and positive at this density: as is well known, the ordered structure has more entropy because particles have more room to wobble about when their mean positions are localized on a lattice\cite{lekkerkerker_colloids_2011}.  

However, within the kinetic approach, the additional entropy cost of localizing {\em distinguishable} particles onto un-swappable lattice sites contains a term $S_{perm} = k_B\ln(N!)$ where $N!$ counts particle permutations. This term must be paid to collapse an accessible phase-space volume in which distinguishable particles can change places, into one where they cannot. This putative entropy cost is supra-extensive ($k_BN\ln N$) and thus for large $N$ outweighs the extensive entropy on formation of the crystal. Thus the kinetic approach to entropy predicts that colloidal crystals are thermodynamically impossible. Yet they are observed every day.

How could a theory get it so wrong? The mistake of the kinetic approach is that it counts the states accessible to only one specific colloidal crystal (i.e., a single permutation of the colloidal positions). But there are $N!$ distinct crystals that might arise, and the kinetic approach underestimates by this enormous factor the probability of finding the system in one or other of them. In the informatic approach, on the other hand, all the states that we as observers deem equivalent are counted as members of the same macrostate; the multiplicity of crystals with different particles occupying different sites then cancels exactly $S_{perm}$, recovering the same result as for indistinguishable particles. Thus we correctly predict that at $\phi = 56\%$ (say) and $N\to\infty$, formation of {\em some crystal or other} occurs with probability one, even though formation of {\em any specified realization of the crystal} occurs with probability zero. The kinetic approach can of course finesse this by adding on a `configurational entropy' to count the number of mutually inaccessible states\cite{glassMa}. But this concedes defeat: the original goal\cite{ma_statistical_1985} was to show that entropy counts only the states the system can actually reach, whereas the configurational term directly adds back all those---within an equivalence class set by the observer's choice of macrostate---that it can't. For thermodynamic purposes, this simply restores the informatic definition of entropy. 

These arguments are not new\cite{eastman_entropy_1933, goldstein_reality_2008, speedy_relations_1999} but deserve to be more widely known (and taught!). 
They illustrate two interconnected points. First, not all reasonable-sounding definitions of entropy for classically distinguishable particles are equivalent: some are right and some are wrong. Second, experiments on colloidal suspensions can resolve with striking clarity what the right definitions are. This adds to the many other ways in which colloidal experiments have clarified basic concepts in classical statistical mechanics\cite{poon_colloids_2004}.

In the rest of this article, we first reiterate the above discussion from a slightly more formal angle (Section~\ref{sec:gibbs}) and then explore similar lines of reasoning in two further areas. In both of these, pitfalls in defining the classical entropy can be illuminated by laboratory experiments, or indeed thought-experiments, involving colloids. The two areas are the so-called `symmetry number' in molecular partition functions (Section~\ref{sec:molecules}); and the paradoxical dependence of rotational entropies on particle masses (Section~\ref{sec:mass}). Finally, in Section~\ref{sec:solvent} we return to the first of the two niggles raised above concerning the success of statistical mechanics: why is it that the entropy of a colloidal system can be computed as though the solvent were replaced by a vacuum? This will also resolve a further paradox involving the role of particle masses in translational, rather than rotational, entropies.
 
\section{Gibbs' paradox and permutation entropy}
\label{sec:gibbs}

Classical statistical mechanics, as formulated by Gibbs\cite{gibbs_collected_1928} rests on the formula $F = -k_BT\ln Z(T,V,N)$ for the Helmholtz free energy, $F$. Here $Z$ is the partition function, which after integrating over momenta can be written for indistinguishable particles as
\begin{equation}
Z = \frac{1}{N!}\frac{1}{\lambda_\textnormal{particle}^{dN}}\int e^{-\beta H({\bf q})}\, d{\bf q}
\label{one}
\end{equation}
In Eq.~\ref{one} the symbol $d$ denotes the dimension of space; $\beta = 1/k_BT$; and $H$ is the classical (configurational) Hamiltonian written as a function of ${\bf q}$, which is a $dN$-dimensional vector of particle coordinates. Finally, $\lambda_\textnormal{particle}$ is a constant, which Gibbs could not calculate, but turns out to be the thermal de Broglie wavelength, 
\begin{equation}
  \label{eq:wavelength}
  \lambda_\textnormal{particle} = \hbar\sqrt{2\pi/m_\textnormal{particle} k_BT},  
\end{equation}
where $m_\textnormal{particle}$ is the mass of the particle of interest, which might later be a single colloidal particle or a molecule. This correspondence is found by re-deriving Eq.~\ref{one} as the semi-classical limit of the quantum partition function: this procedure gives an absolute determination of the integration measure in the classical phase space whose integration over momenta yields the $\lambda$ factors in Eq.~\ref{one}.

By definition, $\lambda$ (which involves $\hbar$) should cancel from any physical observable that involves only classical physics. Its apparent failure to do so---in calculations that are classical, but not quite correct---lies behind some of the paradoxes discussed later. But first let us focus on indistinguishability.

Gibbs was fully aware that for {\em distinguishable} particles the $N!$ divisor in Eq.~\ref{one} is in principle absent, and that this absence destroys thermodynamics as we know it\cite{gibbs_collected_1928}. For instance, without this divisor the entropy of an ideal gas is supra-extensive:
\begin{equation}
S_{IG}(N,V)/k_B \simeq N \ln (V/N\lambda^d) + N \ln N
\end{equation}
Gibbs predated quantum mechanics, but understood (for instance) that all helium atoms are equivalent, hence requiring inclusion of the $N!$ divisor, which restores extensivity by cancelling the second term. Thus the observed extensivity of the thermodynamic entropy in atomic and molecular substances was no surprise to him. 

He did however pose the following question. Suppose we have two types of particle (red and blue) each indistinguishable among themselves, for which $S_{IG}(N_r,N_b,V)/k_B \simeq N_{r}\ln (V/N_{r}\lambda^d)
+N_{b}\ln (V/N_{b}\lambda^d)$. This is not equal to $S_{IG}(N_r+N_b,V)$ for a single species. A paradox now arises if one imagines red and blue particles to smoothly and continuously become merged into a single population of purple particles. How and when does the entropy jump from one formula to another? This is Gibbs' paradox, and one resolution of it is offered by quantum mechanics: you cannot smoothly transmute one chemical species into another, so `red is red and blue is blue and ne'er the twain shall meet'. This resolution asserts that the problem is a mere artefact of classical thinking.

Gibbs himself found a different resolution which, unlike the quantum-mechanical one, makes sense even for classically distinguishable particles such as colloids (which could in effect be made in a continuous range of colours, if one so desired).
He considered how the probabilities of different collections of states would change if the observer chose to `undistinguish' specific subsets of such particles. This act merges previously distinct macrostates, whose probabilities therefore must be added together. Gibbs found that the entropy switches from one form to another at precisely the point where we choose to give up distinguishing red particles from blue. His general analysis of this class of problems remains useful today, for example when considering colloids with a continuous distribution of sizes\cite{warren_combinatorial_1998}. And, in the case where all particles are undistinguished, the final result is simple: the $N!$ divisor reappears in Eq.~\ref{one}, and extensive, single-species thermodynamics is restored. 

Thus, although quantum indistinguishability offers a fast and convenient derivation of the $N!$ divisor in Eq.~\ref{one}, it is quite wrong to believe that this factor defies classical explanation. While many others have noted this fallacy, most recently Frenkel\cite{frenkel_why_2014}, the belief is still widespread (and the supposed breakdown of classical thinking itself often confusingly referred to as the `Gibbs paradox'). The cause of the misunderstanding is perhaps that most of us learn statistical physics from textbooks that rely solely on the quantum mechanical explanation of the $N!$ divisor. In a textbook context  -- where atoms and molecules are typically the main focus (e.g. \cite{kardar_statistical_2013}) -- this represents sound pedagogy; but the result is that many physicists remain unaware of Gibbs' completely classical derivation of the same factor. Indeed it seems that his derivation had been almost entirely forgotten before its excavation by Jaynes about 20 years ago \cite{jaynes_gibbs_1992}. 

\section{Colloidal molecules: entropy and symmetry numbers}
\label{sec:molecules}

\begin{figure}
  \includegraphics{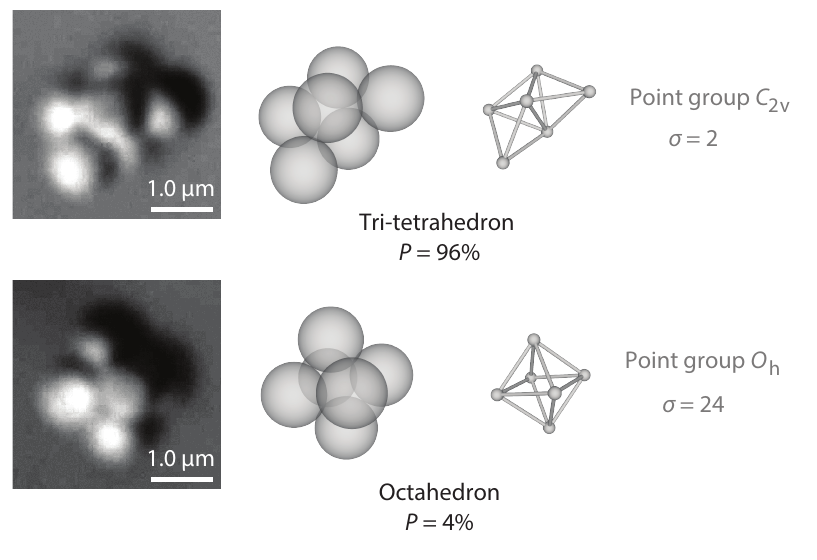}
  \caption{Optical microscope images (left) and structures (middle) of tri-tetrahedral and octahedral colloidal molecules.  Each of these two isomers contains six micrometer-scale polystyrene spheres bound together by depletion forces, which are induced by nanoparticles that cannot be seen in the micrographs.  The depletion attraction, details of which are reported in \cite{meng_free-energy_2010}, has a depth of about 4 $k_BT$ and a range of about 80 nm, much smaller than the size of the polystyrene spheres. Experiments show that in an equilibrium ensemble, the probability of observing a tri-tetrahedron is 96\%, compared to 4\% for the octahedron. The difference in probabilities is dominated by rotational entropy, which is a function of the symmetry number $\sigma$ (right). (Images from Guangnan Meng.)}
  \label{fig:octapoly}
\end{figure}

Perhaps the simplest illustration of Gibbs' argument---and of how our experiment defines the macrostate under observation---comes from experiments on `colloidal molecules': small clusters of micrometer-scale spherical particles that attract one another over a short distance.  The short range of the attraction means that, to a good approximation, the potential energy of a colloidal molecule is proportional to how many `bonds' (pairs of touching particles) it has.  Molecules with different structures but equivalent numbers of particles are called `isomers.'  In three-dimensions, six spheres can form two isomers with 12 bonds, an octahedron and a tri-tetrahedron~\cite{arkus_minimal_2009} (Figure~\ref{fig:octapoly}).  Yet  experiments show that the tri-tetrahedron occurs 24 times as often as its more symmetric isomer, the octahedron~\cite{meng_free-energy_2010}.  

The preponderance of the tri-tetrahedron cannot be explained away by appeal to which structure is more likely to form first, since the isomers can freely interconvert over the timescale of its experiment.  Having ruled out kinetics and potential energy as reasons for the dominance of the tri-tetrahedron, we must conclude that the tri-tetrahedron has a higher entropy than the octahedron.

Where does this entropy difference come from?  We shall show, using the same statistical mechanical formalism used for molecules\cite{mcquarrie_statistical_1976}, that the dominant factor is the difference in symmetry between the two structures.  For simplicity we suppose our classical colloidal `molecule' to be \textit{in vacuo} (the role of a solvent is considered in Section~\ref{sec:solvent}).  In equilibrium, the probability $P$ of observing an isomer is proportional to $Z_\textnormal{mol}e^{-\beta U}$, where $U$ is the potential energy of the isomer (which depends only on its bond number) and $Z_\textnormal{mol}$, its molecular partition function.  To a good approximation, $Z_\textnormal{mol}$ can be factored into translational ($T$), rotational ($R$), and vibrational ($V$) parts: 
\begin{equation}
\label{eq:Zfactored}
Z_\textnormal{mol} = Z_{T}Z_{R}Z_{V}
\end{equation}
The translational part, when converted to a free energy, yields the ideal gas entropy defined above.  It is the same for both isomers, assuming the box containing them is large compared to their dimensions:
\begin{equation}
\label{eq:Ztrans}
 Z_{T}=V/\lambda_\textnormal{molecule}^3.    
\end{equation}
The vibrational part can be calculated by assuming harmonic interactions and taking the product of classical-limit contributions from the normal mode frequencies $\omega_i$: 
\begin{equation}
\label{eq:Zvib}
Z_V=\prod_{i = 1}^{3N-6} \frac{k_BT}{\hbar\omega_i}     
\end{equation}
When this calculation is done, the lower-frequency vibrational modes in the tri-tetrahedron lead to its being favoured by a factor slightly smaller than two. 

We are still off by a factor of approximately 12, which must come from the rotational contribution.  In three dimensions, the classical-limit rotational partition function is
\begin{equation}
\label{eq:Zrot}
Z_{R} = \frac{\pi^{1/2}(2k_BT)^{3/2}}{\hbar^3} \frac{\sqrt{I_1I_2I_3}}{\sigma}
\end{equation}
where $I_{1,2,3}$ are the three principal moments of inertia of the molecule (more on these in Section~\ref{sec:mass}) and $\sigma$ is the symmetry number, defined as the number of ways that a molecule can be rotated and still look the same. The octahedron, a Platonic solid, has a symmetry number of 24, while the tri-tetrahedron, with only one two-fold axis of rotational symmetry, has a symmetry number of 2.  The ratio of 12 between these symmetry numbers, taken together with the factor close to two from vibrational contributions and another close to unity from the moments of inertia, yields the final 24-fold difference in probability.

\begin{figure*}
  \includegraphics{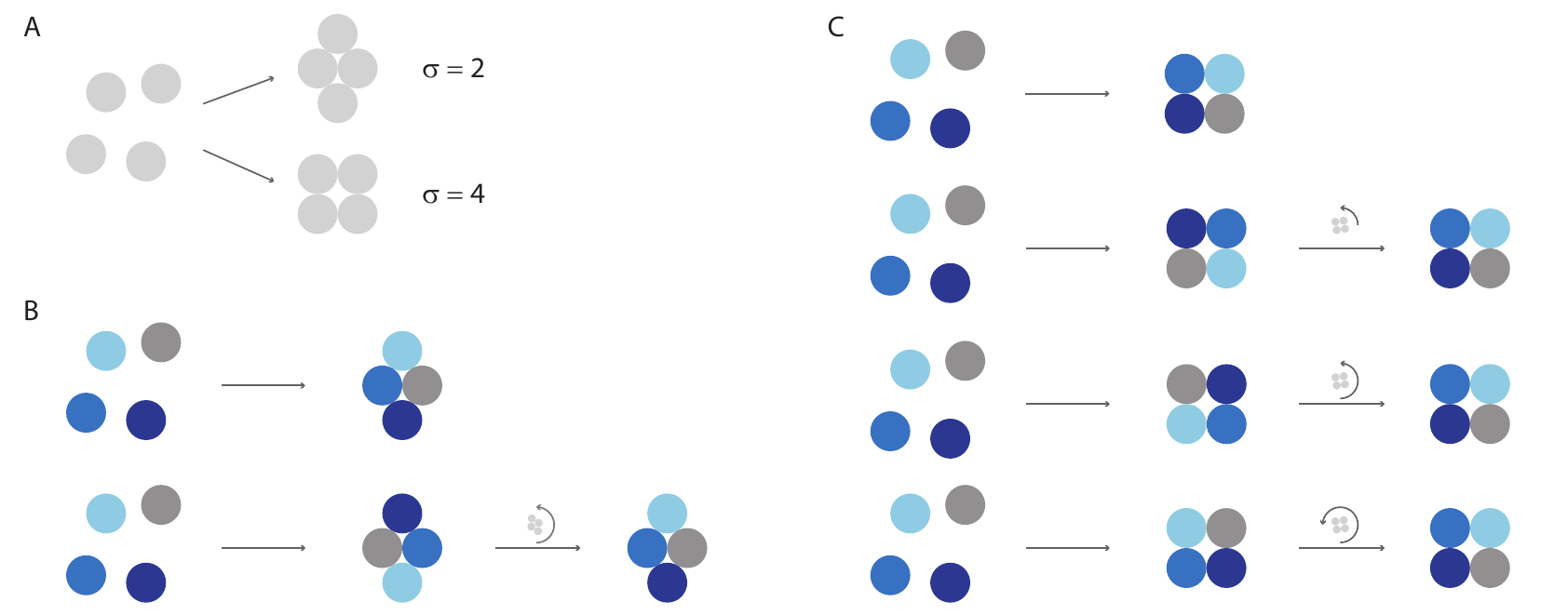}
  \caption{The symmetry number accounts for the relation between permutations and rotations.  A) In this example, four particles come together to form one of two isomers, a rhombus (top) or a square (bottom).  We imagine each cluster to be confined to the two-dimensional plane, so that there is only one rotational axis, which points out of the plane.  Any given orientation of each isomer can be constructed from $4!=24$ permutations of particles.  B) For the rhombus, which has two-fold symmetry in the plane, each configuration of labeled particles (top) is equivalent to a permutation of particles followed by a rotation of $\pi$ radians (bottom).  C) For the square, which has four-fold symmetry in the plane, each configuration can be made in four different ways, corresponding to permutation plus a rotation of
$\pi/2$, $\pi$, or $3\pi/2$ radians.}
  \label{fig:symmetry}
\end{figure*}

Though it is reassuring that the calculation reproduces the experimental results, it sheds little light on how the symmetry of the octahedron could work against it so strongly.  The usual tactic to make sense of such results is to appeal to quantum mechanics.  There the appearance of the symmetry number in the partition function makes perfect sense: the molecular wave function has a symmetry determined by the placement of atoms in the molecule, so that rotations commensurate with this symmetry are fundamentally indistinguishable.  If we did not include the symmetry number, we would overcount the number of rotational microstates.  And indeed this is where many textbooks\cite{mcquarrie_statistical_1976} leave the matter.  

But this explanation is wholly unsatisfying for colloidal molecules, where the wavefunction, were we even able to compute one, would not be symmetric unless each of our particles were composed of exactly the same number and configuration of atoms.  Since that is certainly not the case, and the particles can in principle be distinguished---by small differences in their size, for example---we must seek an alternative explanation.  

To do so, we return to the original derivation of the symmetry number by Ehrenfest and Trkal in 1921\cite{ehrenfest_ableitung_1921, ehrenfest_deduction_1921}, which made no recourse to wavefunctions or quantum mechanics in general.  We illustrate this argument with a simple example inspired by Gilson and Irikura\cite{gilson_symmetry_2010} (see also \cite{calvo_energy_2012}). 

Consider a colloidal molecule with $N=4$ particles that are constrained to two dimensions (Figure~\ref{fig:symmetry}).  There are $4!=24$ possible ways they can be arranged to form a four-particle molecule, for example a rhombus or a square. Writing the partition function in terms of the coordinates of the particles (or `atoms') and integrating out the momenta yields
\begin{equation}
  \label{eq:Zatom}
  Z_\textnormal{atom} = \frac{1}{N!\lambda_\textnormal{atom}^{2N}}\int \exp\left[-\beta U\left(\mathbf{q}_1\ldots\mathbf{q}_n\right)\right]\, d{\mathbf{q}_1\ldots d{\mathbf{q}_n}}.
\end{equation}
The integral extends over all configurations in which the atoms form the molecule of interest.  Because the particles are undistinguished, each configuration can be made in $N!=4!=24$ different ways, and so we divide by $N!$.

Why does the symmetry number appear in $Z_\textnormal{mol}$ (Eq.~\ref{eq:Zfactored}--Eq.~\ref{eq:Zrot}) and not in $Z_\textnormal{atom}$ (Eq.~\ref{eq:Zatom})?  And why does a factor of $1/N!$ appear in $Z_\textnormal{atom}$ and not in $Z_\textnormal{mol}$?  The difference is that $Z_\textnormal{mol}$ is integrated over molecular coordinates.  In writing Eq.~\ref{eq:Zfactored} we assumed that our colloidal molecule is a rigid body that can translate, rotate, and vibrate.  

In molecular coordinates, permutations and rotations are not independent operations but become `entangled' (in the classical sense of that word!).  To see this, we distinguish the particles and assign them colours, as shown in Figure~\ref{fig:symmetry}.  In the distinguished system, the rotational partition function is proportional to the number of permutations times the number of possible orientations of the cluster: $4!\cdot 2\pi$.  But sometimes the same configuration can be made in different ways. For the rhombus, which has two-fold symmetry in the plane, each configuration of labeled particles (Figure~\ref{fig:symmetry}B, top) can also be made by permuting the particles and rotating the cluster by $\pi$ radians (Figure~\ref{fig:symmetry}B, bottom).  The rotational partition function is therefore proportional to $4!\cdot\pi$ or $4!\cdot 2\pi/\sigma$, where $\sigma=2$. For the square, which has four-fold symmetry in the plane, each configuration can be made in four different ways, corresponding to permutation plus rotation of $\pi/2$ radians, or an integer multiple thereof (Figure~\ref{fig:symmetry}C). The rotational partition function is therefore proportional to $4!\cdot \pi/2$ or $4!\cdot 2\pi/\sigma$, where $\sigma=4$.

If, as in the experiment, the particles are undistinguished, we must divide by $N!=4!$ in both cases, leaving a rotational partition function that is proportional to $2\pi/\sigma$.  We now see why the symmetry number appears in $Z_\textnormal{mol}$ and not $Z_\textnormal{atom}$, and vice versa for the $1/N!$ term: both changes result from grouping together the atomic degrees of freedom.  The proportions of the two isomers observed in an equilibrium ensemble will depend inversely on the ratio of their symmetry numbers, as well as on the differences in potential energy, vibrational entropy, and moments of inertia.  (We will see in the next Section that there are some crucial cancellations between the moments of inertia and the vibrational terms.) The same argument applies to the octahedral and tri-tetrahedral colloidal molecules observed in the experiments. 

What if we chose to distinguish the colloidal particles in the experiment, for example by labeling them with different fluorescent colours?  In that case it would be natural to consider each different arrangement of coloured particles to be a different molecule. There are 30 arrangements for the octahedron ($6!/\sigma$, where $\sigma=24$), and 360 for the tri-tetrahedron ($\sigma=2$).  If we were to count how often we saw each of these isomers, we would find that each of the 360 tri-tetrahedral isomers would occur twice as often as each of the 30 octahedral isomers, because of the ratios of moments of inertia and vibrational frequencies.  It is only when we lump these macrostates together---by ignoring the distinctions between particles---that we obtain a result that depends on symmetry numbers.

It is no coincidence that this argument is similar to that made by Gibbs in resolving the paradox now named after him.  The aim of Ehrenfest and Trkal was in fact to to understand the origin of the $N!$ divisor in the canonical partition function, the very same factor that we discussed in Section~\ref{sec:gibbs}.  The only way to understand how the entropy of a gas depends on the number of molecules, Ehrenfest and Trkal argued, is to consider a situation where the number of molecules $N$ can vary in a \emph{reversible process}.  Within the canonical ensemble, $N$ is fixed when the system contains only a single species, and we cannot add more molecules reversibly.  But when there are multiple species of molecules in the same system that can interconvert, their numbers $N_i$ may vary reversibly. In this case, factors of $N_i!$ appear in the partition function without any appeal to quantum mechanical argument.  The symmetry number appears as a side-note in this thought experiment.  It is a historical irony that most modern textbooks ascribe to it a quantum origin, given that Ehrenfest and Trkal aimed to show exactly the opposite (see \cite{gilson_symmetry_2010} and \cite{van_kampen_gibbs_1984} for elegant and modern reformulations of their arguments).  

\begin{figure}
  \includegraphics{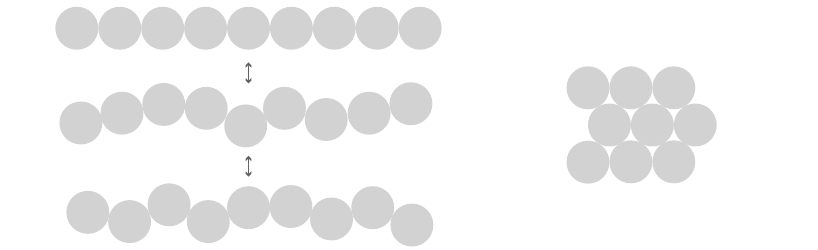}
  \caption{Left, a flexible linear molecule that can fluctuate between high-symmetry (top) and low-symmetry (middle, bottom) conformations without changing its bond network. Right, a symmetric isomer.  If our experiment does not distinguish between the various states on the left and only counts the  abundance of any conformation on the left versus the one on the right, we must assign a symmetry number of 2 to the flexible molecule, even though it spends most of its time in asymmetric conformations.}
  \label{fig:flexible}
\end{figure}

The broader point here is that our statistical mechanical models must be consistent with the types of observation we make, since those observations \emph{define} the macrostate to be modeled.  Consider, for example, a flexible colloidal molecule that can fluctuate between symmetric and asymmetric conformations without changing its bond topology (Figure~\ref{fig:flexible}).  Does such a molecule have a symmetry number, given that it spends most of the time in asymmetric conformations?  The answer depends on what we observe.  If our experiment measures only the concentration of this molecule relative to other isomers, then the symmetry number must appear in our calculation, because it accounts for how identical conformations can be reached by permuting particles followed by either rotation or flexing\cite{gilson_symmetry_2010}. This example shows that the symmetry number is more generally understood as a property of the topology of the bond network of a molecule rather than its geometry.  This may seem like a deep statement, but it is really a definition that follows from what experiments typically measure (and what they do not). Put differently, we have \emph{chosen} to treat all three states of the chain in Fig.~\ref{fig:flexible} as states of the same molecule, not different molecules, and we must calculate the symmetry number accordingly.

\section{Colloidal entropies are mass-independent}
\label{sec:mass}

As already discussed, the relative abundance of different colloidal `molecules' should be proportional to the partition function for each type. The partition function of a molecule includes the symmetry number as just described, but also factors counting the number of translational, vibrational and rotational states. In a quantum context, it is conventional to consider each of these factors separately -- in part because in many molecular systems at room temperature rotations are almost classical, while vibrations are not.

When used to address classical entropies, this artificial separation leads to another paradox. (Recall that a paradox is an apparent contradiction that, on close inspection, disappears.) Let us consider the rotational contribution to its partition function which in three dimensions is Eq.~\ref{eq:Zrot}. For the simplest case of a diatomic molecule, deriving this result is a standard undergraduate exercise: one examines the small $\beta$ limit of the quantum partition function $Z = \sum_n (2n+1)\exp(-\beta E_n)$ where $E_n = \hbar^2 n(n+1)/2I$ is the rotational energy and $(2n+1)$ the number of distinct quantum states of that energy. Eq.~\ref{eq:Zrot} can be derived by generalizing this quantum calculation, or by a purely classical integral over the canonical coordinates for a rigid rotor.

The paradox is as follows. Imagine we have two species of colloids of the same size and interactions but with two different particle masses. For instance, the two types could comprise inner spheres of gold and of aluminium, each coated with an outer layer of polystyrene so as to create the same final size and surface chemistry. For simplicity ignore gravity (assume the experiment to be done in the space station), and consider `triatomic' linear molecules that contain one gold and two aluminium particles. Then according to Eq.~\ref{eq:Zrot}, the gold particle is more likely to be found in either one of the edge positions than in the middle position. More generally, given two species of particles, placing the heavy ones at the periphery of the molecule increases the moment of inertia and hence the partition function. According to Eq.~\ref{eq:Zrot} this should directly influence the abundances of these different configurations, favouring those with heavy particles outermost. In atomic physics this would be called an `isotope effect'.

Can this really be true? A thought-experiment says no: colloids explore their configuration space by over-damped Brownian motion in which inertia is negligible, so how can the inertial mass possibly control the abundances of different cluster geometries? The paradox was raised in a commentary by one of us\cite{cates_self-assembly_2012}, with the tentative suggestion that Eq.~\ref{eq:Zrot} might fail for particles embedded in a molecular solvent rather than a vacuum (see Section~\ref{sec:solvent} below). But in fact the resolution of this paradox is much simpler than that\cite{herschbach_molecular_1959,boresch_jacobian_1996}. The `colloidal isotope effect' is a mirage, caused by artificially separating classical rotations from vibrations and translations. 

It is true that for a molecule in vacuo the full classical partition function is usually approximated by the factorized form, Eq.~\ref{eq:Zfactored}. But what matters is whether this object, not just one of its factors, supports an isotope effect. It doesn't, as we can see by sidestepping the above factorization and using instead a more fundamental `interacting atom' representation:
\begin{eqnarray}
Z &=& \int \exp\left[-\beta(H_T+H_R+H_V)\right]\, d{\bf p} \, d{\bf q} \nonumber\\
 &=& \int \exp\left[-\beta\left(\sum p_i^2/2m_i + U({\bf q})\right)\right] \, d{\bf p} \, d{\bf q} \nonumber\\
 &=& Z_KZ_C \label{fullpart}
\end{eqnarray}
Here $Z_K$ and $Z_C$ are the kinetic and configurational contributions arising from the ${\bf p}$ and ${\bf q}$ integrals, and $U({\bf q})$ is the potential part of the Hamiltonian. Because
the ${\bf p}$ and ${\bf q}$ integrals are separable, and because we assumed by hypothesis that $U({\bf q})$ is mass-independent, the probability of observing different configurations of the particles is fixed solely by their interactions, and not by their masses. Eq.~\ref{fullpart} is written down for point particles; in principle extended objects such as colloidal spheres will have a further factor from the rotational kinetic energy of the spheres themselves, but this is again factorable and makes no difference to the argument.

How does this outcome square with the `molecular' representation, Eq.~\ref{eq:Zfactored}? For a classical molecule of $N$ atoms---assuming harmonic vibrational states and no vibration-rotation coupling---the molecular partition function (in $d=3$) follows from Eqs.~\ref{eq:Zfactored}--\ref{eq:Zrot}: 
\begin{equation}
Z_\textnormal{mol} = \frac{V}{\lambda_\textnormal{molecule}^3} \frac{\pi^{1/2}(2k_BT)^{3/2}}{\hbar^3} \frac{\sqrt{I_1I_2I_3}}{\sigma}\prod_{i = 1}^{3N-6} \frac{k_BT}{\hbar\omega_i}
\end{equation}
where the $\omega_i$ are the angular frequencies of the vibrational states. The particle mass ratios enter not only the $I$'s but also the $\omega$'s; and an exact cancellation between these two factors -- while far from obvious in this representation -- is completely guaranteed by Eq.~\ref{fullpart}\cite{herschbach_molecular_1959}. 

\section{Integrating out the solvent}
\label{sec:solvent}

Let us finally return to the first of the two niggles mentioned in the introduction concerning the success of statistical mechanics: the fact that colloids move through a solvent, not empty space. To see what the problem is, consider the quantum statistical mechanics of a dilute atomic gas. The ideal gas contribution to the entropy (after allowing for the $N!$ divisor in Eq.~\ref{one}) is
\begin{equation}
S_{IG} = Nk_B\ln\frac{V}{N\lambda_\textnormal{particle}^d} + \left( \frac{d}{2}+1 \right)Nk_B
\end{equation}
As is well known, the factors of $\lambda$  generally cancel out in the entropy {\em differences} that control phase behaviour -- specifically, when comparing states with the same number of translational degrees of freedom, such as a liquid and a gas of the same chemical species. Suppose however that our particles can dimerize, forming a rigid bond of binding energy $\Delta$. Then one finds that the equilibrium constant relating the concentrations of dimers ($D$) and monomers ($M$) obeys
\begin{equation}
K \equiv  \frac{c_D}{c_M^2} = 2^{d/2}\lambda_M^d \exp\left[\beta\Delta\right]. \label{K}
\end{equation}
This is a simple application of the Boltzmann distribution; it shows that where processes are present that can change the number of degrees of freedom, quantum contributions to the ideal-gas entropy play a role. 

Let's now address colloids in a solvent. A good starting point is the full semi-classical partition function for all particles present, in which we retain the $Nd$ colloid coordinates ${\bf q}$ but integrate out the solvent ones ${\bf Q}$. Integrating also over all momenta as usual, we have 
\begin{equation}
Z({\bf p},{\bf q}) = \frac{1}{M!\lambda_\textnormal{solvent}^{Md}}\int\,\exp\left[-\beta H({\bf p},{\bf q},{\bf Q})\right]\, d{\bf Q}
\equiv \exp\left[-\beta U({\bf q})\right] \label{PMF}
\end{equation}
where $M$ the number of solvent molecules\cite{frenkel_introduction_2006}.
The second form {\em defines} the effective interaction potential of the colloids $U({\bf q})$. Eq.~\ref{PMF} is, of course, another example of strategically discarding information (the solvent coordinates) that under other circumstances we might have chosen to retain.

So long as we can determine a good approximation for $U({\bf q})$, then the solvent has indeed gone away; our colloids might as well be in vacuo, so long as this effective interaction replaces the one they would really have had there. The effective interaction potential can include solvation forces and all manner of other solvent-dependent interactions; nonetheless, all that we need to know about the solvent is encoded in $U({\bf q})$. 

The paradox lies in the continuation of this argument. The colloid free energy is now $F(T,V,N) = -k_BT\ln Z$ where
\begin{equation}
Z =\frac{1}{N!\lambda_\textnormal{colloid}^{Nd}}\int \,\exp\left[-\beta U({\bf q})\right]\, d{\bf q} \label{reduced}
\end{equation}
This still contains the thermal de Broglie wavelength of the colloids, $\lambda_\textnormal{colloid}$. It might appear then that if our potential $U({\bf q})$ has a short range deep attraction, conventionally represented by a bond energy $\Delta$, Eq.~\ref{K} will still apply. If so, the relative populations of colloidal dimers (for instance in our mixture of gold-cored and aluminium-cored colloids with identical interactions) depends explicitly on their particle masses. This seems very wrong, for the same reasons as discussed in the previous section when comparing relative abundances of different cluster shapes. What is more, the actual value of $K$ seemingly involves $\hbar$ even though we ought to be dealing with purely classical physics here\cite{frenkel_introduction_2006}. 

One resolution sometimes offered is that, as the solvent gets integrated out, the effective phase-space measure for the colloids -- which roughly defines how far you have to move a colloid before it counts as being in a `different' configuration, becomes set not by $\lambda$ but by a coarse-grained length-scale $\hat\lambda$ related to the solvent diameter\cite{cates_self-assembly_2012}. This is, for example, how the world might appear to a lattice modeller where the solvent size appears to set a natural discretization scale. 
The classical quantity $\hat\lambda$ is then postulated ad-hoc to replace $\lambda$ in Eq.~\ref{reduced}, and hence in Eq.~\ref{K} and all similar results for formation of supra-dimeric clusters.

But in fact this is a specious argument, for reasons spelled out below. Instead the astute reader will have noticed that the paradox is based on the same flawed reasoning as the one discussed in the previous section: we have treated some degrees of freedom as classical, but not all. Specifically, we assumed that the binding energy $\Delta$ of a colloidal dimer was connected with formation of a fixed bond -- ignoring the fact that, classically, all bonds have a vibrational entropy.  Referring again to Eq.~\ref{fullpart}, it is once again clear that -- because masses only enter the momentum integral in the partition function, which is separable -- all masses must cancel, as must $\hbar$, once the full set of classical degrees of freedom is treated on an equal footing.

The way this happens for colloidal bonding is easiest seen for the case of a square well potential, of depth $\Delta$ and range $\ell$. We treat for simplicity the case of $d = 1$, for which rotations do not enter. However small $\ell$ may be, in the classical limit there is a finite entropy of confinement for the relative coordinate between the two particles in our dimer; and for the square well this is simply $S_{bond} = k_B\ln(\ell/\lambda)$. The equilibrium constant is now (using Eq.~\ref{K} for $d=1$) $K = \lambda\sqrt{2}\exp[\beta \Delta + S_{bond}]= \ell\sqrt{2} \exp[\beta\Delta]$. Thus the equilibrium constant is independent of quantum mechanics and there is, as promised, no isotope effect. 

We now see why the replacement of $\lambda$ with $\hat\lambda$ did not offer a true resolution to the paradox. If the physical behaviour is classical, none of the actual results can depend on $\lambda$; so replacing $\lambda$ with $\hat\lambda$ can make no difference {\em unless the calculations have been wrongly executed}. As the above example shows, the length that replaces $\lambda$ in (say) the dimerization constant $K$ of Eq.~\ref{K} is not some ad-hoc coarse-graining scale $\hat\lambda$: it is instead a precisely defined length that can be calculated directly from the effective interaction $U({\bf q})$. And in our chosen example, it is simply the range of the bonding interaction, $\ell$. 

In summary, for classical systems such as colloids, the absolute phase-space measure as set by quantum mechanics must cancel from any observable physical quantity. This remains just as true after integrating out the solvent (to get an effective interaction $U({\bf q})$) as it was before doing so. It is therefore never wrong to use for the colloid partition function the measure set by the semi-classical limit of quantum mechanics, in which factors of the thermal de Broglie wavelength $\lambda$ appear, Eq.~\ref{reduced}. But if such factors fail to cancel in any predicted observable, this is because a mistake has been made. The most likely pitfall is to inadvertently treat some of the classical degrees of freedom as frozen. (Indeed, freezing a classical degree of freedom carries an infinite classical entropy cost, which is cut off only by quantum mechanics.) Because the phase-space measure must cancel, replacing $\lambda$ with a classical coarse-graining length such as a solvent size, while it may disguise an incorrect calculation, cannot change a correct one.

\if{
\begin{figure*}[tbp]
\caption{}
\label{fig1}
\end{figure*}
}\fi

\section{Conclusion}

A century of quantum mechanics has had a profound influence on the way we think about statistical physics. The paradoxes and ambiguities we have discussed all arise from assumptions and explanations based on the quantum viewpoint.  Sometimes these assumptions are so deeply embedded in the formalism that we forget they are there. For example, the vibrations of most molecules are quantized even at room temperature; we would get nonsense if we interpreted molecular spectra using a classical formalism.  Consequently, a textbook might not even consider the classical limit of the vibrational entropy.  But only by taking this limit can we understand why the thermal wavelength---an absurdly small lengthscale for a macroscopic object like a colloidal particle---cancels out in the equilibrium constant for dimerization of such objects.

In other cases, assumptions carried over from quantum mechanics are mostly harmless, but not if they cause one to forget that the same results can also be justified classically. For example, there is little danger in assuming that colloidal particles are indistinguishable, since most experiments choose to leave them undistinguished, which gives equivalent results in the classical limit. Danger does lie in arguing that because colloids are distinguishable in principle, theory can be improved by treating them as distinguished in statistical mechanics. By defining macrostates far too narrowly, that line of thinking leads to predictions that are not borne out by experiments: colloids would never crystallize, and symmetric colloidal molecules would have the same rotational entropy as asymmetric ones.

Experiments on colloids have helped to resolve other paradoxes as well.  The oldest and most famous of these is Maxwell's demon. In the 1980s Landauer and Bennett `exorcised' the demon by showing that erasing classical information carries an entropy cost~\cite{bennett_demons_1987}. But for decades afterward it was not unusual to find articles~\cite{maddox_maxwells_2002} claiming that the demon must produce entropy when it makes a measurement.  As Bennett noted~\cite{bennett_information_1998}, the tendency to focus on the entropy cost of measurement, rather than of erasure, may `have been a side effect of the great success of quantum mechanics.'  The question has been settled by recent experiments on colloidal particles, which show that the cost of erasing information agrees with Landauer's prediction~\cite{berut_experimental_2012}.

These experiments and the others we have mentioned all support the informatic view of entropy.  The main argument against this view is that it introduces subjectivity into the definition of the entropy.  This is true, but we see it as a strength, not a weakness.  We use the term `subjective' in the same sense as Jaynes~\cite{rosenkrantz_e._1983} did: `depending on the observer.' For classical systems such as colloids, the observer has a choice of what to observe and what to ignore; for example, she can treat each different arrangement of six fluorescent particles as a different molecule, or she can ignore the colours and keep track of only the bonds.  The macrostate therefore depends on the observer's choice.  So, too, does the entropy, which is found by counting the number of microstates that make up this macrostate. It is possible to define the entropy non-subjectively---for example, by insisting that colloids always be treated as distinguishable---but we have seen that such a definition makes it difficult to reconcile theory with experiment.

A subjective entropy might appear to create new problems.  After all, the heat transferred in a reversible process is proportional to the change in entropy.  If the entropy depends on the observer, how can a measurement of the heat be consistent from observer to observer?  We note first that experiments can measure only entropy differences, not absolute entropies.  Each observer is free to add an arbitrary (subjective) constant to the entropy, and even if each chooses a different constant, the measurements are unaffected.  But what if the heat were measured directly, for example by measuring the motion of the particles with an optical microscope? Would the heat flow really depend on the choices of the observer? Perhaps surprisingly, the answer is yes, because the observer can choose which degrees of freedom to measure and which to ignore.  In any reversible change, the energy change associated with degrees of freedom that are measured is called work; the remainder is called heat. The first law of thermodynamics (which states that the sum of heat and work is the change of internal energy) then directly requires that the heat flow depends on which degrees of freedom are retained as state variables, and which are ignored. But if each observer makes the same choice, the heat measured will be the same from observer to observer---and in that sense it remains an objective quantity.

Thus we would argue that the informatic view is the simplest way to interpret experiments on colloids.  As long as we clearly specify our choices, all paradoxes vanish.  The beauty of colloidal systems is that the particles are fundamentally classical and distinguishable in principle. This allow us consciously to make choices about the macrostate in more than one way, and to observe and compare the results. In so doing, we gain a better understanding of Gibbs' arguments and the effects of entropy in the macroscopic world.

\section{Acknowledgements}

This work was funded in part by EPSRC grant no. EP/J007404 and NSF grant no. DMR-1306410. MEC holds a Royal Society Research Professorship. We thank
Jean-Philippe Bouchaud,
Michael Brenner,
Alasdair Bruce,
Paul Chaikin,
Joshua Deutsch,
Alexander Grosberg,
Guangnan Meng,
Onuttom Narayan,
David Pine,
Rebecca Perry,
Jorg Schmalian,
Frans Spaepen,
David Wales,
Patrick Warren,
Tom Witten,
and
Mathieu Wyart
for discussions on colloidal entropy during the past several years. VNM thanks the organizers and students of the 2013 UMass Summer School on Soft Solids and Complex Fluids and the Center for Soft Matter Research at NYU for providing the opportunity to study these problems.

\bibliographystyle{rsc}
\providecommand*{\mcitethebibliography}{\thebibliography}
\csname @ifundefined\endcsname{endmcitethebibliography}
{\let\endmcitethebibliography\endthebibliography}{}

\end{document}